\documentclass[preprint,preprintnumbers,aps,amsmath,amssymb,floatfix,showpacs,superscriptaddress,prl]{revtex4}

\usepackage{graphicx} 
\usepackage{bm} 
\usepackage{ulem} 
\usepackage{pifont}

\begin{document}

\title{Creation of New Lasing Modes with Spatially Nonuniform Gain}
\author{Jonathan Andreasen}
\altaffiliation[Also at]{
  Department of Physics and Astronomy, Northwestern University,
  Evanston, IL 60208, USA
}
\affiliation{
  Department of Applied Physics, Yale University,
  New Haven, CT 06520, USA
}
\author{Hui Cao}
\affiliation{
  Department of Applied Physics, Yale University,
  New Haven, CT 06520, USA
}
\affiliation{
  Department of Physics, Yale University,
  New Haven, CT 06520, USA
}
\date{\today}
\begin{abstract}
  We report on the creation of new lasing modes with spatially nonuniform 
  distributions of optical gain in a one-dimensional random structure.
  It is demonstrated numerically that even without gain saturation and mode 
  competition, the spatial nonuniformity of gain can cause dramatic and 
  complicated changes of lasing modes.
  New modes appear with frequencies in between those of the lasing modes with 
  uniform gain.
  We examine some new lasing modes in detail and find they exhibit high output 
  directionality.
  Our results show that the random lasing properties may be modified 
  significantly without changing the underlying structures. 
\end{abstract}
% 42.55.Zz, Random lasers
% 42.60.-v, Laser optical systems: design and operation
% 42.55.Ah, General laser theory 
\maketitle 

%-------------------------------------------------------------------------------
A conventional laser consists of a resonant cavity and amplifying material. 
The lasing modes have a nearly one-to-one correspondence with the resonant 
modes of the cold cavity \cite{siegbook}. 
Except for a slight frequency pulling, the lasing properties are usually 
determined by the cavities. 
Thus, cavity design is essential to obtain desirable lasing frequencies or 
output directionalities. 
The available lasing modes are typically fixed once the cavity is made. 
Finer control over lasing properties can be obtained, for example, by carefully 
placing the gain medium in a cavity to reduce the lasing threshold 
\cite{horo92} or using specific pumping profiles to select lasing modes with 
desirable properties \cite{gmachl98,fuku02,chern03,hent09}.
However, once the laser cavity is made, it is very difficult to obtain new 
lasing modes that have no correspondence to the resonant modes of the cold 
cavity if nonlinearity is negligible. 

A random laser is made of disordered media and the lasing modes are 
determined by the random distribution of refractive index. 
Because of the randomness, it is difficult to intentionally produce lasing 
modes with desirable properties. 
To have more control over random laser properties, the structures themselves 
may be adjusted by selecting the scatterer size 
\cite{wu04,vannestepre05,gott08,garc09,garc09am} and separation 
\cite{souk04,lagen07},
changing the scattering structure with temperature \cite{wiersma,lawandy}
or electric field \cite{wiersmaPRL}, or creating defects \cite{fuji09}.
For random lasers operating in the localization region, spatially 
non-overlapping modes may be selected for lasing through local pumping of the 
random system \cite{sebbah02}. 
In the case of diffusive random lasers, far above the lasing threshold, 
nonlinear interaction between the light field and the gain medium alters the 
lasing modes \cite{tureciSci}. 
Without gain nonlinearity, local pumping and absorption in the unpumped region 
can also change the lasing modes significantly \cite{yamilovOL} because the 
system size is effectively reduced.
Recent experiments \cite{polson,wu06} and numerical studies \cite{wu07} show 
that even without absorption in the unpumped region, the spatial 
characteristics of lasing modes may vary with local pumping. 
In this case, the lasing modes still correspond to the resonant modes of the 
passive system. 
However, spatial inhomogeneity in the refractive index can introduce a linear
coupling of resonant modes mediated by the polarization of gain medium
\cite{deych05PRL}.

In this Letter, we demonstrate that new lasing modes can be created by 
nonuniform distributions of optical gain in one-dimensional (1D) random 
systems without absorption and nonlinearity.  
These new lasing modes do not correspond to the modes of the passive system or 
any lasing modes in the presence of uniform gain.
They typically exist for specific gain distributions and disappear as the 
profile is further altered. 
They can lase independently of other lasing modes when gain saturation is 
taken into account. 
The new lasing modes appear at various frequencies for many different gain 
distributions and can have highly directional output. 
These findings may offer an easy and fast way of dramatically changing the 
random laser properties without modifying the underlying structures.  

We consider a 1D random system composed of $N=161$ layers.
Dielectric material with index of refraction $n_1=1.05$ separated by air gaps
($n_2=1$) resulting in a spatially modulated index of refraction $n(x)$.
The system is randomized by specifying different thicknesses for each of the
layers as $d_{1,2} = \left<d_{1,2}\right>(1+\eta\zeta)$, where
$\left<d_1\right>=0.1$ $\mu$m and $\left<d_2\right>=0.2$ $\mu$m are the average
thicknesses of the layers, $\eta = 0.9$ represents the degree of randomness, 
and $\zeta$ is a random number in (-1,1).
The length of the random structure $L$ is normalized to 
$\left<L\right>=24.1$ $\mu$m.
The index of refraction outside of the random media is $n_0 = 1$.
The above parameters give a localization length of $\xi \approx 240$ $\mu$m at 
a vacuum wavelength $\lambda = 600$ nm, which is the wavelength of interest in 
this work.

The transfer matrix (TM) method developed in \cite{wu07} is used to simulate 
lasing modes at the threshold with linear gain.
A real wavenumber $k = 2\pi/\lambda$ describes the lasing frequency.
Propagation of the electric field through the 1D structure is calculated via 
the $2\times 2$ transfer matrix $M$. 
The boundary conditions at the lasing threshold with only emission out of the
system require $M_{22}=0$.
Linear gain is simulated by appending an imaginary part to the index of 
refraction $\tilde{n}(x) = n(x) + in_if_E(x)$, where $n_i < 0$.
We neglect the change of the real part of the refractive index $n(x)$ in the 
presence of gain.
Spatial nonuniformity of gain is implemented by multiplying the imaginary part 
by a step function $f_E(x)=\Theta(-x+l_G)$, where $x=0$ is the left edge of 
the random structure and $x=l_G$ is the location of the gain edge on the right 
side.

%-------------------------------------------------------------------------------
\begin{figure}
  \includegraphics[width=4.2cm]{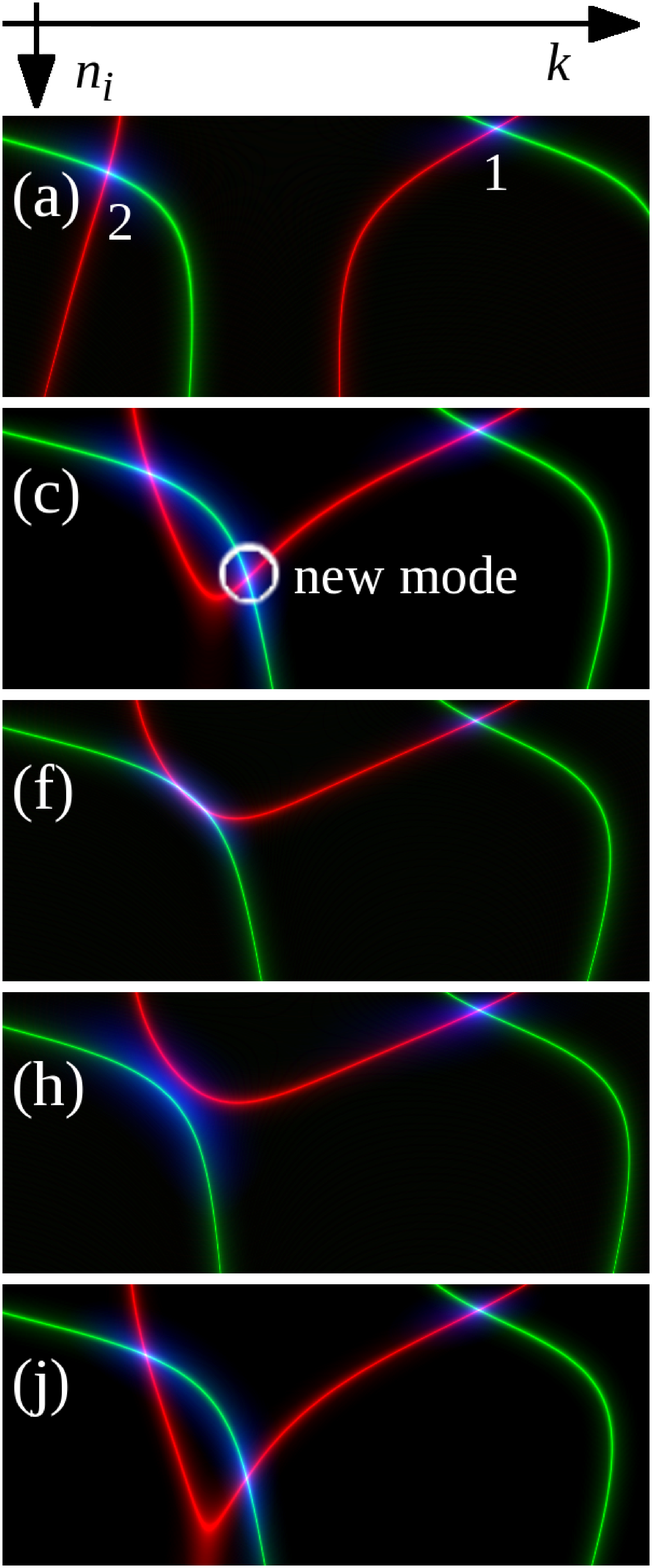}
  \includegraphics[width=4.2cm]{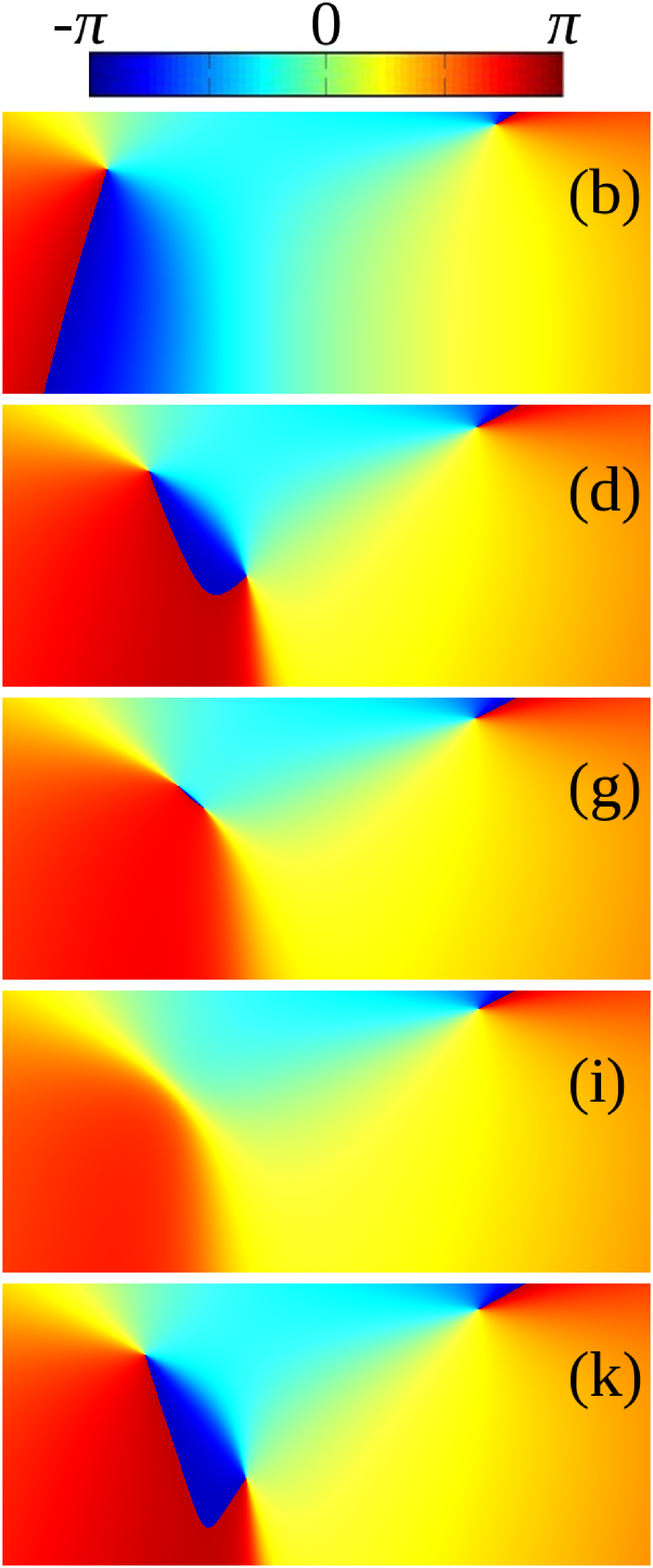}
  \caption{\label{fig:fig1} (Color online)
    Left column: Real (green) and Imaginary (red) zero lines of $M_{22}$ in 
    the $(k, n_i)$ plane. 
    Right column: Phase $\theta$ of $M_{22}$ in the $(k, n_i)$ plane. 
    The horizontal axis is real wavenumber $k$ in the range 
    $10.48$ $\mu$m$^{-1}$ and $10.64$ $\mu$m$^{-1}$. 
    The vertical axis is the imaginary part of refractive index $n_i$ in the 
    range -0.0130 and -0.0326.
    The length $l_G$ of the gain region is from top to bottom: 
    (a-b) $14.961$ $\mu$m, (c-d) $14.559$ $\mu$m, (f-g) $14.523$ $\mu$m, 
    (h-i) $14.472$ $\mu$m, (j-k) $14.295$ $\mu$m.
  }
\end{figure}

%-------------------------------------------------------------------------------
Lasing frequencies and thresholds (gain required to induce lasing) are located 
by determining which values of $k$ and $n_i$, respectively, satisfy $M_{22}=0$.
$\operatorname{Re}[M_{22}]=0$ ($\operatorname{Im}[M_{22}]=0$) forms real 
(imaginary) ``zero lines'' in the ($k$, $n_i$) plane.
The crossing of a real and imaginary zero line results in $M_{22}=0$ at that
location, thus pinpointing a solution.
We visualize these zero lines in Fig. \ref{fig:fig1} by plotting 
$-\log_{10}\left|\operatorname{Re}M_{22}\right|$ and
$-\log_{10}\left|\operatorname{Im}M_{22}\right|$ together and using image 
processing techniques to enhance the contrast.
We monitor the changes of such zero lines as the gain edge is moved gradually 
from $l_G=L$ (uniform gain) within the wavelength range 
500 nm $< \lambda <$ 750 nm.
Appearances of new lasing modes with frequencies in between those of the 
lasing modes with uniform gain are observed.
Figure \ref{fig:fig1} concentrates on a smaller frequency range with 
Fig. \ref{fig:fig1}(a) showing two of the lasing modes (marked mode 1 and 
mode 2) resulting from the crossing of zero lines for $l_G=14.961$ $\mu$m.
Mode 1 at $\lambda_1=592.6$ nm has a lower lasing threshold than mode 2 at 
$\lambda_2=598.0$ nm.
They are the only two lasing modes found within this frequency range and their 
frequencies are almost the same as those of the corresponding resonant modes 
of the passive system. 
As $l_G$ is decreased, the lasing modes do not shift much in frequency. 
However, when $l_G=14.559$ $\mu$m, the zero lines are joined as shown in 
Fig. \ref{fig:fig1}(c).  
A new lasing mode, encircled in white, appears in between modes 1 and 2.
The spatial intensity distribution of the new lasing mode differs from those 
of modes 1 and 2. 
As $l_G$ decreases further, the zero lines forming mode 2 and the new mode 
pull apart.
This causes the solutions to approach each other in the ($k$, $n_i$) plane, 
seen in Fig. \ref{fig:fig1}(f), until becoming identical.
The zero lines then separate resulting in the disappearance of mode 2 and the 
new mode as evidenced by Fig. \ref{fig:fig1}(h).
The lines cross again for $l_G=14.295$ $\mu$m and the solutions reappear and 
move away from each other in the ($k$, $n_i$) plane in Fig. \ref{fig:fig1}(j).

%-------------------------------------------------------------------------------
Verification of lasing mode solutions is provided by the phase of $M_{22}$, 
calculated as 
$\theta=\operatorname{atan2}(\operatorname{Im}[M_{22}],\operatorname{Re}[M_{22}])$.
Locations of vanishing $M_{22}$ give rise to phase singularities.
The phase change around a closed path surrounding a singularity is referred to 
as topological charge \cite{halperin,ZhangJOSAA}.
Two phase singularities are seen in Fig. \ref{fig:fig1}(b) at the same 
locations as the zero line crossings in Fig. \ref{fig:fig1}(a).
This verifies the authenticity of the lasing mode solutions.
The phase singularity at the location of the new mode in the ($k$, $n_i$) plane
[Fig. \ref{fig:fig1}(d)] confirms that it is a genuine lasing mode in the
presence of linear gain.
The phase singularity associated with the new mode is of opposite charge to 
the existing ones.
As $l_G$ is reduced, two oppositely charged phase singularities move closer
[Fig. \ref{fig:fig1}(g)] and eventually annihilate each other at 
$l_G=14.295$ $\mu$m [Fig. \ref{fig:fig1}(i)]. 
As already mentioned, this process reverses itself and the two lasing modes 
reappear in Fig. \ref{fig:fig1}(k).

%-------------------------------------------------------------------------------
For a more thorough study of the new lasing modes and confirmation of their 
existence in the presence of gain saturation, we switch to a more realistic 
gain model including nonlinearity. 
The Bloch equations for the density of states of two-level atoms \cite{ziol95} 
are solved together with the Maxwell's equations with the finite-difference 
time-domain method \cite{tafl05}.
The phenomenological decay times due to the excited state's lifetime $T_1$ and 
decoherence $T_2$ are included. 
The gain spectral width is given by 
$\Delta\omega_a=1/T_1+2/T_2$ \cite{siegbook}. 
We also include incoherent pumping of atoms. 
The rate of atoms being pumped from the ground state to the excited state is 
proportional to the ground state population, and the proportional coefficient 
$P_r$ is called the pumping rate. 
The resulting Maxwell-Bloch (MB) equations are solved numerically with the 
spatial grid step $\Delta x = 1.0$ nm and the temporal step 
$\Delta t = 3.3\times 10^{-18}$ s.
The atomic density $N_{atom}/V = 1.8 \times 10^{13}$ cm$^{-3}$.
Nonuniform gain is simulated by having the two-level atoms only in the region 
$0 \le x \le l_G$.

By setting the atomic transition wavelength $\lambda_a$ to coincide with the   
wavelength of mode 1, 2 or the new mode and using a narrow gain spectrum, we 
are able to investigate the three lasing modes separately.
$\Delta\omega_a$ is chosen to be less than the mode spacing to ensure single 
mode lasing (at smaller pumping rates).
At $l_G=14.295$ $\mu$m, the wavelength difference between mode 2 and the new 
mode, which is smaller than that between mode 1 and the new mode, is 
$\lambda_2-\lambda_{nm}=1.4$ nm. 
We set $T_1 = 1.0 \times 10^{-12}$ s and $T_2 = 0.73 \times 10^{-12}$ s so 
that the gain spectral width in terms of wavelength is 
$\Delta\lambda_a =0.71$ nm.
Initially all atoms are in their ground state and the system is excited by a 
Gaussian-sinusoidal pulse with center wavelength $\lambda_0=\lambda_a$ and 
spectral width $\Delta\lambda_0=\Delta\lambda_a$.
When the pumping rate $P_r$ is above a threshold value, the electromagnetic 
fields build up inside the system until a steady state is reached.

%-------------------------------------------------------------------------------
\begin{figure}
  \includegraphics[width=8.5cm]{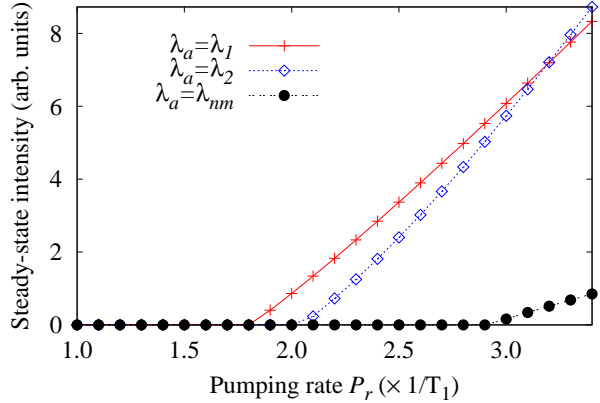}
  \caption{\label{fig:fig2} (Color online)
    Steady-state output intensity vs. pumping rate $P_r$ from Maxwell-Bloch 
    simulations with different gain spectra.
    The wavelength of the atomic transition $\lambda_a$ is equal to 
    $\lambda_1$ (red crosses), 
    $\lambda_2$ (blue open diamonds), and
    $\lambda_{nm}$ (black circles). 
  }
\end{figure}

%-------------------------------------------------------------------------------
Figure \ref{fig:fig2} shows the steady-state output intensity with 
$\lambda_a=\lambda_1$, $\lambda_2$, or $\lambda_{nm}$ as $P_r$ is varied.
$P_r=1/T_1$ corresponds to the transparency point, namely, the excited state 
population of atoms is equal to that of the ground state. 
The lasing threshold pumping rate for mode 1 is reached first at 
$P_r=1.9/T_1$, then mode 2 at $P_r=2.1/T_1$ and the new mode at $P_r=3.0/T_1$.
These thresholds agree qualitatively with the values of the TM calculation.

%-------------------------------------------------------------------------------
\begin{figure}
  \includegraphics[width=8.5cm]{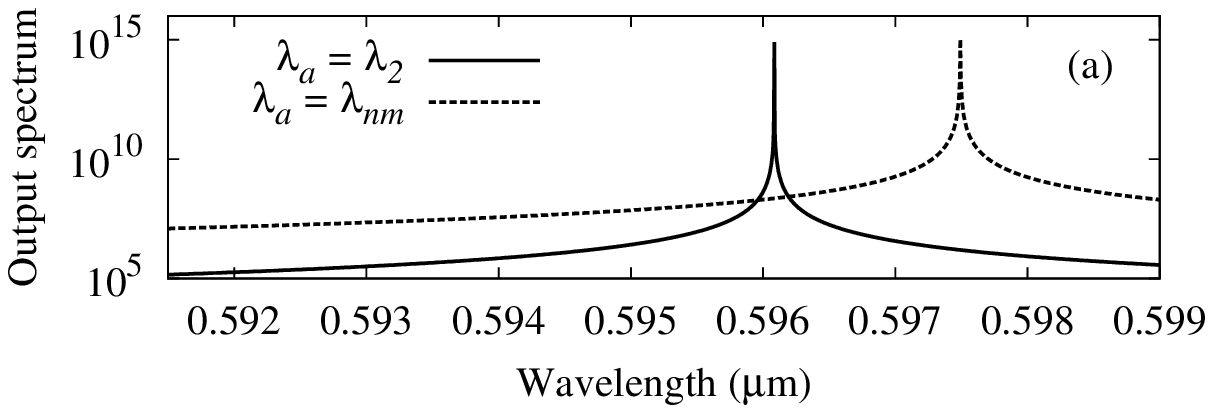}
  \includegraphics[width=8.5cm]{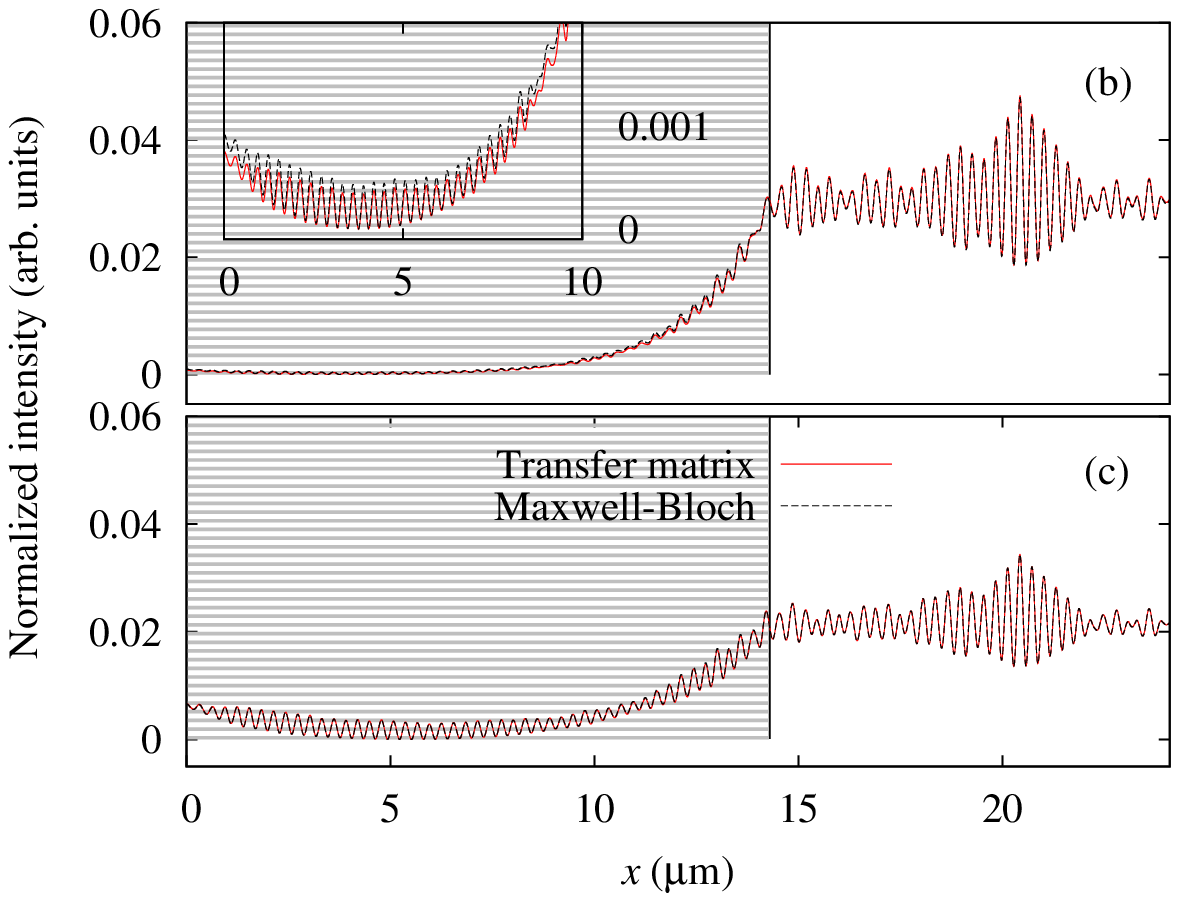}
  \caption{\label{fig:fig3} (Color online)
    (a) Emission spectra taken from two Maxwell-Bloch simulations with 
    $\lambda_a$ equal to $\lambda_{nm}$ (solid line) and 
    $\lambda_2$ (dashed line).
    In both cases, $\Delta\lambda_a = 0.71$ nm, $l_G=14.295$ $\mu$m. 
    $P_r=3.0/T_1$ for the solid curve and $2.1/T_1$ for the dashed curve. 
    The emission intensity reaches a steady state by a simulation time of 
    $t=300\times 10^{-12}$ s.
    %The spectral resolution $\Delta\lambda$ is given by the temporal 
    %integration length of the Fourier transformation. 
    The spectra are taken from $1400\times 10^{-12}$ s to 
    $1700\times 10^{-12}$ s. %meaning $\Delta\lambda = 0.0045$ nm.
    The MB intensity distribution $|\phi_{MB}(x)|^2$ (black dashed line) is
    compared to the TM intensity distribution $|\phi_{TM}(x)|^2$ 
    (red solid line) for the new lasing mode (b) and mode 2 (c).
    The inset in (b) is an expansion of the curves for 0 $\mu$m $<x<10$ $\mu$m.
  }
\end{figure}

%-------------------------------------------------------------------------------
When $\lambda_a = \lambda_{nm}$, the first lasing mode is the new mode, 
instead of mode 1 or 2. 
Figure \ref{fig:fig3}(a) shows the output emission spectrum just above the 
lasing threshold at $P_r = 3.0/T_1$. 
It consists of a single lasing mode with the wavelength equal to that of the 
new mode calculated with the TM method.
The spatial intensity distribution obtained from the Maxwell-Bloch (MB) 
calculation is compared to that from the TM calculation in 
Fig. \ref{fig:fig3}(b). 
The MB distribution $|\phi_{MB}(x)|^2$ is found by integrating the intensity 
over one optical period.
It is then normalized to the TM distribution $|\phi_{TM}(x)|^2$ as
$\int_0^L |\phi_{MB}(x)|^2 dx = \int_0^L |\phi_{TM}(x)|^2 dx$.
The two intensity distributions are almost identical. 
The average percent difference between them is 7.77\%. 
This result indicates the nonlinear effect due to gain saturation is small 
when the pumping rate is just above the lasing threshold. 
When the peak of the gain spectrum is shifted from $\lambda_{nm}$ to 
$\lambda_1$ or $\lambda_2$, the first lasing mode is switched to mode 1 or 2. 
Figure \ref{fig:fig3}(c) plots the spatial intensity distribution of mode 2 
obtained by the MB calculation with $\lambda_a = \lambda_2$ and 
$P_r = 2.1/T_1$ as well as that obtained by the TM calculation. 
The two distributions are almost the same and they are different from the 
distribution with uniform gain. 
Comparing Fig. \ref{fig:fig3}(b) to (c), we see the spatial intensity 
distribution of the new lasing mode differs significantly from that of mode 2 
within the gain region. 
Outside the gain region the two distributions are not much different because 
their wavelengths are very close. 

When optical gain is located on the left side of the structure, we observe 
that the intensity distributions of new lasing modes are heavily concentrated 
on the right side of the gain region. 
This makes the emission intensity through the right boundary of the random 
system much larger than that through the left boundary. 
We calculate the the ratio of right to left output flux 
$S\equiv |\phi(x=L)|^2/|\phi(x=0)|^2$. 
For the new  lasing mode in Fig. \ref{fig:fig3}(b) $S_{nm}=40$, indicating the 
laser output is mostly to the right. 
As a comparison, $S_{1}=1.1$ for mode 1 and $S_{2}=3.3$ for mode 2. 
Thus, the new lasing mode has much more directional output than modes 1 and 2. 

%-------------------------------------------------------------------------------
Because of the excellent agreement found between the MB and TM calculations, 
we conclude that new lasing modes do appear in random lasers with spatially 
nonuniform distributions of optical gain.
Typically, as in the case studied here, they are sensitive to the spatial gain 
distribution and disappear if the distribution is altered slightly. 
These new lasing modes offer more control of random laser performance as their 
properties such as frequency and output directionality can be quite different 
from those of existing lasing modes. 
Moreover, the properties of new lasing modes can be easily altered by varying 
the spatial profile of the pump beam, without modifying the random structures. 

%-------------------------------------------------------------------------------
The authors thank Christian Vanneste, Patrick Sebbah, Li Ge, A. Douglas Stone, 
Jan Wiersig, and Dimitry Savin for stimulating discussions, 
and acknowledge support from the Yale Faculty of Arts and Sciences
HPC facility and staff.
This work was supported partly by the National Science Foundation under Grant 
Nos. DMR-0814025 and DMR-0808937.

\end{document}